Title: **GHz Dielectric Measurement of Powdery TiO**

Authors: Robert Tempke*, º, Christina Wildfire+, º, Dushyant Shekhawat+, Terence Musho*, +, z

Abstract: This study experimentally determines the dielectric properties of a powdery titanium (II) oxide (TiO) material within the microwave range (0.1-13.5 GHz). The properties were determined using a coaxial airline method using a TiO/paraffin mixture at several loading fractions. A permittivity of 60 for volume fraction below 30% and 100 for volume fraction above were measured.

* West Virginia University, Department of Mechanical and Aerospace Engineering, WV 26506, USA
+ Department of Energy - National Energy Technology Laboratory, Morgantown, WV 26507,
º Oak Ridge Institute of Science and Education
z Corresponding Author - email: tdmusho@mail.wvu.edu.

1. Introduction

Experimental characterization of powder materials interactions with electromagnetic waves is a critical factor in the development of new devices and process at GHz frequencies. It is found that the dielectric properties of powder materials are often different from the bulk material properties. This stems from the geometric localization of the dipoles opposed to the continuum dipole interactions. Particle geometry and volume fraction are all influenced by the electrical (dipole) continuity of the sample and thus the polarizability or dielectric properties of the sample. This study focuses primarily on experimentally determining the dielectric constant of titanium (II) oxide (TiO) in the range 0.1-13.5GHz. TiO was selected because of its high dielectric constant that is potentially tunable by controlling the oxygen stoichiometry [1, 2]. TiO is a metallic conductor, a superconductor at 5.5K [3], and is abundant [4]. TiO is highly insoluble and has good thermal stability [1], which makes it favorable as an additive or coating in capacitor dielectrics [5]. It is hypothesized that this material may exhibit constant dielectric properties over the frequency range of interest, making it a promising material for GHz devices and catalyst for modular chemical reactors.

This study focuses on providing statistically significant dielectric measurements of TiO using a paraffin-powder composite method. The methodology of setting the powder into a paraffin matrix was chosen for this study because of the electrically conductive nature of the powder and to understand the influence of volume loading on the dielectric properties. A coaxial transmission line measurement was used to determine the dielectric properties of the composite mixtures. The mixtures' dielectric properties were subsequently analyzed to predict the dielectric constant of the constitutive TiO using an empirical mixture equation [6].

2. Experimental

The powdery material used as the composite filler for testing was titanium (II) oxide (Alfa Aesar, 99.9%) and the paraffin was fully refined paraffin wax. The composites plugs were created using a paraffin matrix with the previously mentioned powder at various volume loadings. Powder volume loadings were selected to be 5, 10, 20, and 30 percent. Powders were combined with solid paraffin pellets in a beaker and heated to 70°C using a water bath. Once the paraffin was fully melted, the powder was thoroughly mixed using mechanical agitation and cast into a 3D printed inverted test cell. This inverted test cell was surrounded by a cylindrical sleeve with an inner diameter corresponding to the outer diameter of the test cells airline. The composite was then allowed to cool to room temperature and cut to the desired length.

To ensure a high degree of accuracy, this study created six composite plugs for each volume loading of the powders and tested them individually. Each plug was tested from 0.1-13.5 GHz and the scattering parameters (S-parameters) were measured at 51 equally spaced intervals. The material properties were calculated from the scattering parameters using the Nicholson-Ross-Weir polynomial method [7]. Dielectric measurements were made using a 7-mm diameter coaxial airline (HP model no. 85051-60010) connected to a Keysight N5231A PNA-L microwave network analyzer. The dielectric constant results from the measured paraffin-TiO composites were analyzed using the well-known Looyenga mixing rule shown in Equation 1. This equation was chosen based on the guidelines laid out for powder-paraffin composites extraction calculations [8]. The values of the various volume loadings and dielectric constants of the composite and paraffin were substituted into the equation and the dielectric constant of TiO was calculated. This was done for each plug at every volume loading.

$$\varepsilon_{mix}^{1/3} = V_m \varepsilon_m^{1/3} + V_p \varepsilon_p^{1/3} \qquad 1$$

Where $\varepsilon_{mix}$, $\varepsilon_m$, and $\varepsilon_p$ stand for the dielectric constant of the mixture composite, matrix material (paraffin) respectively, and powdery material. $V_p$ is the volume loading of the powder and $V_m$ is the volume loading of the paraffin. The phase composition of the as-obtained powder was confirmed using XRD. The diffraction peaks were indexed to the structure of TiO (JCPDS # 01-073-8760) with no other phases present as shown in Figure 1. It can be seen from Figure 1 that since Ti is a 3d transition element it forms a monoxide in the crystallite form of NaCl structure.

To achieve a smooth surface on the particles and to avoid discontinuities in particle sizes, ball milling in conjunction with a Brunauer–Emmett–Teller (BET) method was employed to find the surface area of the particles. After ball milling, for 6 hours the sample had a BET surface area of 4.30 m²/g. With single point adsorption total pore volume of pores less than 1286 Å diameter at P/Po = 0.98 and single point desorption total pore volume of pores less than 1091 Å diameter at P/Po = 0.98 of 0.018 cm³/g and 0.019 cm³/g respectively. This pore size and deemed sufficiently small to allow heterogenous mixing of the paraffin, TiO solution.

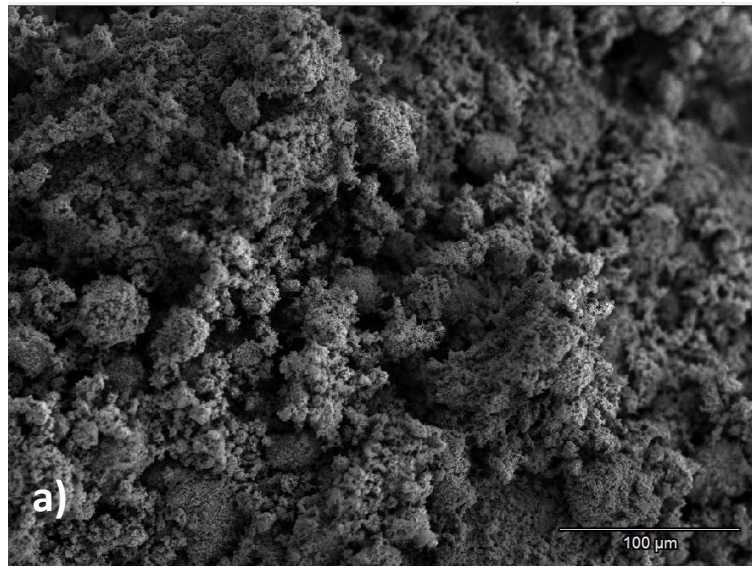
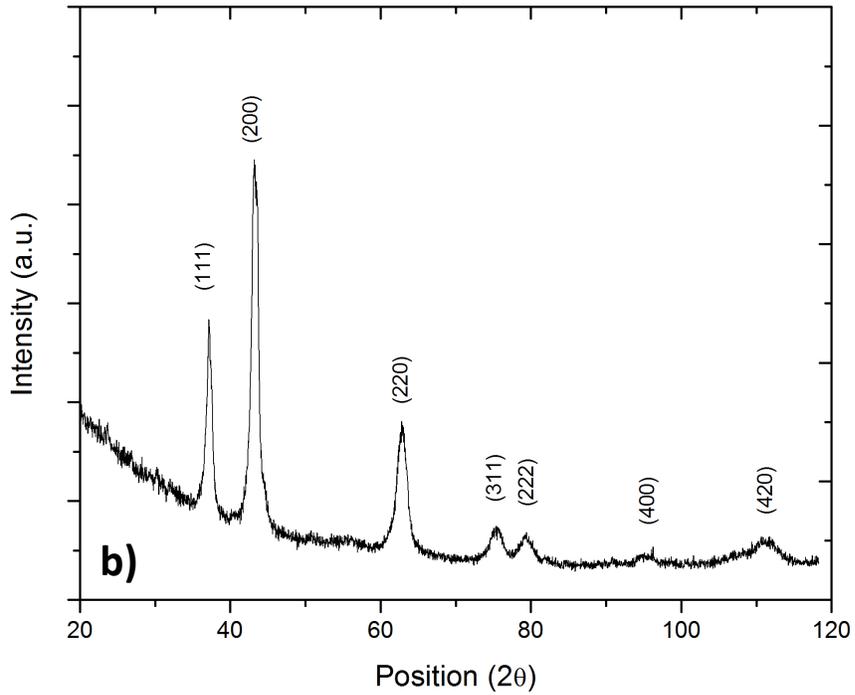

Figure 1: Subfigure a) is SEM image of TiO powder. Subfigure b) is a XRD measure of pure TiO powder, which conforms to that of a NaCl structure. XRD reference JCPDS #01-073-8760

3. Results

A single phase homogenous paraffin plug was tested in the frequency range of interest and found to have a constant dielectric constant of 2.3 across all frequencies. The frequency independent nature of the paraffin makes it an ideal matrix for testing frequency dependent powders. Figure 2 illustrates dielectric constants of powder-paraffin mixture at different volume loadings. The scatter plot points in Figure 2 are representative of 51 measured values across the entire frequency range of 0.1-13.5 GHz. The lines in Figure 2 provide an estimate of the dielectric constant of the mixture as a function of volume loading using the Looyenga mixing rule. Since the unknown in the equation is the constitutive dielectric constant of TiO, solving the equation using different estimated values for TiO will provide insight into the bounds of your dielectric constant. Based on which lines bound the points one can make an estimate of the dielectric constant of TiO and see any if there is any variability related to frequency dependence.

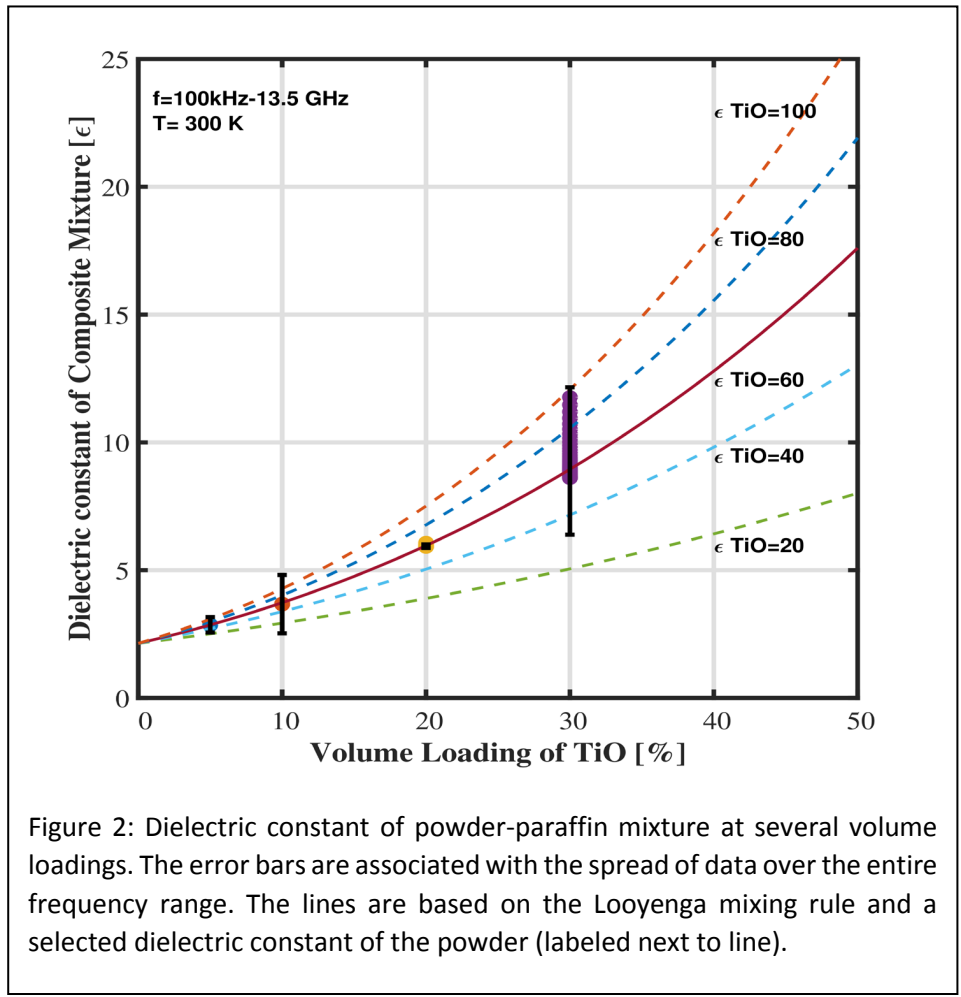

Figure 2: Dielectric constant of powder-paraffin mixture at several volume loadings. The error bars are associated with the spread of data over the entire frequency range. The lines are based on the Looyenga mixing rule and a selected dielectric constant of the powder (labeled next to line).

The large spread in the data shown in Figure 2 is synonymous with the random error associated with dielectric testing, but more so with the frequency dependence of the powder's dielectric constant. Using the Looyenga mixing rule solved for $\varepsilon_m$ the constitutive dielectric constant of TiO can be accurately mapped against frequency. Figure 3 shows the calculated dielectric constant of the constitutive TiO powder as a function of frequency. For volume loadings transitioning from 20% to 30%, there is a notable frequency dependence. This frequency dependence is reliant on a percolation threshold. At this percolation, the powder particles are situated close enough to each other to induce neighboring interaction of dipoles. It is reasonable to believe that the spacing and thus the volume fraction provides a method of controlling the frequency dependence of the powder. At volume fractions below 20%, the dielectric constant was approximately 60. While for a volume fraction of 20% the experimental points are between 60 and 90. Using the experimentally determined dielectric constant of the mixture the constitutive dielectric properties of TiO as a function of frequency was determined using the Looyenga Mixing Rule. Figure 3 is a plot of the frequency dependence of the powder between 0.1 and 13 GHz. It is found that at lower volume fractions (<20%) that the dielectric constant is approximately 60 with no distinguishable frequency dependent. At volume fraction of 20 and 30%, there is an increasing frequency dependence.

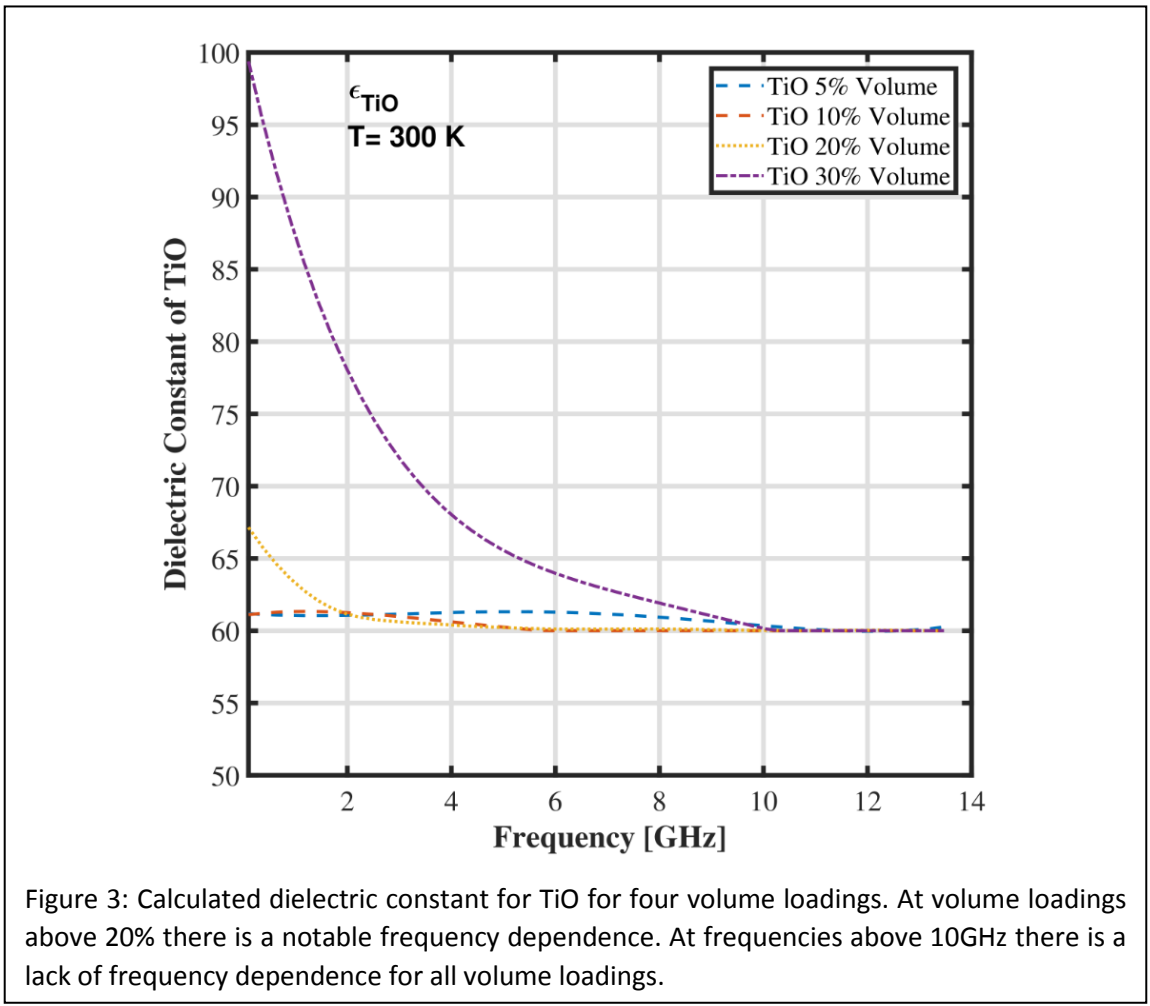

Figure 3: Calculated dielectric constant for TiO for four volume loadings. At volume loadings above 20% there is a notable frequency dependence. At frequencies above 10GHz there is a lack of frequency dependence for all volume loadings.

4. Conclusions

This study experimentally determines the dielectric constant of a TiO powdery material within the frequency range of 0.1-13 GHz. This is accomplished by back calculating the dielectric constant from the composite mixture via the Looyenga mixing rule for two-phase homogenous composites. At the higher volume fractions above 10%, a frequency dependence was shown in the composite and therefore the constitutive powdery material. This frequency dependent was presumably due to a result of the neighboring interaction of the dipoles of the powder particles. Bulk TiO is reasoned to be frequency dependent due to the increased continuity of the induced electric dipoles across the material. However, in powder form, TiO at low volume fractions below 10% was determined to have a dielectric constant of approximately 60. At a volume fraction of 30% the dielectric constant varied between 60 and 100. It is reasoned that since the 30% volume loading mixture has achieve the percolation threshold, it represents the true dielectric constant of bulk TiO.

5. Acknowledgements

This research was supported in part by an appointment to the National Energy Technology Laboratory Research Participation Program, sponsored by the U.S. Department of Energy and administered by the Oak Ridge Institute for Science and Education.